# Pandemic Drugs at Pandemic Speed: Infrastructure for Accelerating COVID-19 Drug Discovery with Hybrid Machine Learning- and Physics-based Simulations on High Performance Computers

Agastya P. Bhati[1], Shunzhou Wan[1], Dario Alfè[2,3], Austin R. Clyde[4], Mathis Bode[5], Li Tan[6], Mikhail Titov[7], Andre Merzky[7], Matteo Turilli[7], Shantenu Jha[6,7], Roger R. Highfield[8], Walter Rocchia[9], Nicola Scafuri[9], Sauro Succi[10], Dieter Kranzlmüller[11], Gerald Mathias[11], David Wifling[11], Yann Donon[12], Alberto Di Meglio[12], Sofia Vallecorsa[12], Heng Ma[13], Anda Trifan[13], Arvind Ramanathan[13], Tom Brettin[14], Alexander Partin[13], Fangfang Xia[13], Xiaotan Duan[4], Rick Stevens[14], Peter V. Coveney[1,15*]

[1]Centre for Computational Science, University College London, Gordon Street, London, WC1H 0AJ, UK
[2]Department of Earth Sciences, London Centre for Nanotechnology and Thomas Young Centre at University College London, University College London, Gower Street, London WC1E 6BT, UK
[3]Dipartimento di Fisica Ettore Pancini, Università di Napoli Federico II, Monte Sant'Angelo, I-80126 Napoli, Italy
[4]Department of Computer Science, University of Chicago
[5]Institute for Combustion Technology, RWTH Aachen University, Aachen 52056, Germany
[6]Brookhaven National Laboratory, Upton NY, 11973
[7]Rutgers, the State University of New Jersey, Piscataway, NJ 08854, USA
[8]Science Museum, Exhibition Road, London, SW7 2DD, UK
[9]Concept Lab, Italian Institute of Technology, Via Melen, Genova, Italy
[10]Center for Life Nanosciences at La Sapienza, Italian Institute of Technology, viale Regina Elena, Roma, Italy
[11]Leibniz Supercomputing Centre (LRZ) of the Bavarian Academy of Sciences and Humanities, Boltzmannstraße 1, 85748, Garching bei München, Germany
[12]OpenLab, CERN, Geneva, Switzerland
[13]Data Science and Learning Division, Argonne National Laboratory, Lemont, IL 60439, USA
[14]Computing, Environment and Life Sciences Directorate, Argonne National Laboratory, Lemont, IL, 60439, USA
[15]Institute for Informatics, Science Park 904, University of Amsterdam, 1098 XH Amsterdam, The Netherlands

Email: p.v.coveney@ucl.ac.uk

**Keywords:** Machine Learning, Artificial Intelligence, Novel Drug Design, Molecular Dynamics, Free Energy Predictions

**Abstract**

The race to meet the challenges of the global pandemic has served as a reminder that the existing drug discovery process is expensive, inefficient and slow. There is a major bottleneck screening the vast number of potential small molecules to shortlist lead compounds for antiviral drug development. New opportunities to accelerate drug discovery lie at the interface between machine learning methods, in this case developed for linear accelerators, and physics-based methods. The two *in silico* methods, each have their own advantages and limitations which, interestingly, complement each other. Here, we present an innovative infrastructural development that combines both approaches to accelerate drug discovery. The scale of the potential resulting workflow is such that it is dependent on supercomputing to achieve extremely high throughput. We have demonstrated the viability of this workflow for the study of inhibitors for four COVID-19 target proteins and our ability to perform the required large-scale calculations to identify lead antiviral compounds through repurposing on a variety of supercomputers.

1. **Introduction**

The COVID-19 pandemic has shaken the world, and the scale and rapidity of the crisis have also challenged existing methods of doing research, not least the current drug design process, which takes about 10 years and $1-3 billion to develop a single marketable drug molecule [1,2]. The disease is caused by the novel severe acute respiratory syndrome coronavirus 2 (SARS-CoV-2), a member of the coronavirus family, which was first identified in the mid-1960s at the Common Cold Unit in Wiltshire, England [3]. Discovering how to combat the pandemic rests on understanding recent outbreaks, such as severe acute respiratory syndrome coronavirus (SARS-CoV), which has the most closely-related genome, and middle east respiratory syndrome coronavirus (MERS-CoV), and taking advantage of the explosion of research in 2020 on various aspects of SARS-CoV-2 biology, from transmission to life cycle. Based on this research, notably experimentally-derived structures for the various viral target proteins, several drug repositioning and drug designing studies have been conducted using *in silico* computer-based modelling technologies [4–6]. However, the identification of conclusive drug molecules has been hampered by the huge chemical space that needs to be explored.

Because of the vast number of potential ligands (ranging from a few hundred million to billions), it is clearly not possible to synthesize them in wet labs, nor is it desirable given that most of them are not going to bind with SARS-CoV-2 proteins at all. This is where *in silico* methods can play an important role in screening the binding affinity of ligands with SARS-CoV-2 proteins to identify and rank potential drug candidates.

There is an increasingly large number of *in silico* methods available to screen candidate ligands. The two most popular categories are physics-based (PB) techniques including molecular dynamics (MD) based methods, and machine learning (ML) techniques. However, both have inevitable limitations and, even after months of research there is a disappointing lack of potential drug candidates for COVID-19 given that so many lives are at stake. There is an urgent need to accelerate the current drug design process and the work presented here is a step in the direction.

PB techniques involve *ab initio* as well as semi-empirical methods which are fully or partially derived from firm theoretical foundations [7–10]. For example, MD is a popular approach for conformational sampling which is derived from Newtonian equations of motion and the concepts of statistical thermodynamics. MD-based free energy calculation methods have been widely applied for predicting protein-ligand binding affinities and subject to extensive experimental validation [11-22]. There are many such free energy methods, some "approximate"; others more "accurate".

In the last decade or so, ensemble simulation based methods have been proposed which overcome the issue of variability in predictions from MD-based methods due to their extreme sensitivity to simulation initial conditions [13,19–23] which leads to chaotic behaviour, and non-Gaussian statistics [24,25]. In particular, two methods named Enhanced Sampling of Molecular Dynamics with Approximation of Continuum Solvent (ESMACS) [13,14,17,26] and Thermodynamic Integration with Enhanced Sampling (TIES) [19–22] have been shown to deliver accurate, precise and reproducible binding affinity predictions within a few hours. Their excellent scalability allows them to calculate binding affinities for a large number of protein-ligand complexes in parallel, utilizing the large size and multiple cores of current supercomputers.

Another important factor affecting the reliability of results is the extent of conformational sampling achieved by MD simulations. Thus, several enhanced methods have also been developed to better sample the phase space [27]. However, even such enhanced sampling is prone to variability in results due to extreme sensitivity to initial conditions. Once again, ensemble simulations are required to control uncertainties in predictions [20–22].

However, these *in silico* methods are computationally demanding and are unable to explore the extensive chemical space relevant for drug molecule generation. To focus the hunt, they require extensive consultation with chemists to suggest structural features or specific functional groups that may improve a ligand's interaction with the target protein, based on the chemical environment of the binding pocket. Drawing on human intelligence and insights takes time and slows the process of drug discovery by delaying the pipeline of candidate ligands to wet labs for testing. Even if this step is accelerated, another bottleneck in drug design looms because there is a limit to the number of compounds that can be studied experimentally.

To overcome the limitations of PB methods, ML methods can be employed. Prediction of binding affinities using a deep neural network has been an active area of research over the past few years. ML represents a set of techniques that rely on inferring complex relationships from big data and applications include diverse fields such as robotics, gaming, language processing and chemoinformatics. Some examples include classifying kinase conformations [28], predicting antimicrobial resistance [29], modelling quantitative structure-activity relationships [30], and predicting contact maps in protein folding [31], with AlphaFold making important progress in protein structure prediction [32].

In the field of drug discovery, ML, specifically deep learning (DL), allows us to generate novel drug-like molecules by sampling a significant subset of the chemical space of relevance. DL techniques are computationally much cheaper and enable quick turnaround of results which allows millions to billions of compounds to be handled [33]. Recent developments in DL allow generation of novel drug-like molecules *in silico* by sampling a large fraction of the chemical space of relevance (estimated to be about $10^{68}$ compounds). However, the accuracy of ML/DL methods is very much dependent on the training data. Their predictive capability can be improved by providing them with reliable data and by curating them with theoretical understanding [34], neither of which may always be available. This restricts their applicability in the drug discovery domain.

ML and PB methods have their own advantages and limitations. Fortunately, their strengths and weaknesses complement each other and so it makes sense to couple them in drug discovery. In the past few years, several attempts have been made to create synergies between PB and ML methods in order to get favourable outcomes. A major application has been to enhance sampling in MD simulations which includes learning of optimal biasing potentials, optimal collective variables or free energy surfaces [35-42]. Examples are also available for approaches that involve deriving MD-based descriptors that can be used to train ML models for predictions of solvation, hydration and/or transfer free energies [43-45]. Studies have shown that the accuracy of alchemical free energy predictions may be improved by "correcting" them through ML-based post-processing [46-47]. In addition, it has been reported that the prediction of ligand activity/affinity against a target can be achieved with a combination of MD and ML [48-51]. Recently, a method combining DL and MD for generation of antimicrobial peptides has been reported where DL methods were used to generate 90,000 peptide sequences which underwent *in silico* screening to finally obtain 20 sequences for experimental validation [52].

ML/DL techniques can be employed to augment human intelligence (HI) with artificial intelligence (AI) for exploring the large chemical space to predict "useful" ligand molecules. This substantially speeds up the process of ligand discovery. On the other hand, reliable PB free energy methods can rank the ligands on the basis of their binding affinities and ground the simulations on theoretical understanding. These binding affinities can then be fed back into the DL algorithm to augment its knowledge base and hone predictions. Such a combination can be an effective tool for drug design and can prove useful in prospective drug design projects. Robust predictive mechanistic models are of particular value for constraining ML when dealing with sparse data, exploring vast design spaces to seek correlations and, most importantly, testing if correlations are causal [53].

It is well accepted that drug targets can undergo significant conformational changes during their biological activity. Some of these changes may involve large-scale rearrangements, such as a domain motion over a hinge region, while some others may be more limited in size, such as the short-lived opening of a mostly hydrophobic cryptic site. The interesting point is that they can involve targetable structures that might otherwise remain hidden to experimental structural determination. Although the physics-based models, such as MD simulations, can explore conformational space to some extent, they can hardly achieve ergodicity, resulting in some of the potential new target structures remaining hidden. Here DL approaches are envisaged to explore whether a short stretch of a MD trajectory may exhibit the hallmarks of potentially biologically relevant structural transitions, even though such transitions are *not* observed in the trajectory itself.

Not only will exploitation of AI ensure that the best use is made of medicinal chemists for drug discovery, it also helps counter chemists' bias during exploration of the chemical space. Carefully trained DL algorithms may be expected to reach regions of the extensive chemical space that may remain untouched by humans.

In this work, we present a novel *in silico* method for drug design by coupling ML with PB methods. We bring together several methods into a coherent scientific workflow - some of which are already being applied in drug discovery while others are relatively new to the field and yet to be adopted. Rather than performing only blind ML/DL, we couple them with accurate PB methods to make them "smarter". Potential candidates are selected from the output of a DL algorithm and they are scored using PB methods to calculate binding free energies. This information is then fed back to the DL algorithm to refine its predictive capability. This loop proceeds iteratively involving a variety of PB scoring methods with increasing level of accuracies at each step ensuring that the DL algorithm gets progressively more "intelligent". As described above, several methods employing a constructive combination of ML and PB methods have been reported in the past few years. However, the pipeline described in this article is unique in several ways. We attempt to generate ligand structures with improved binding potency towards a given target protein using an iterative loop with both upward and downward exchange of information at each step – this, we believe, has not been attempted before. We posit that our innovative integration of PB and ML-based methods can substantially reduce the throughput time for exploring a huge chemical space and improve the efficacy of exploration of chemical libraries for lead discovery. It is worth mentioning here that since the success of lead molecules identified at pre-clinical stages is heavily dependent on several factors like membrane permeability, toxicity, water solubility, etc., drug repurposing provides another avenue for quick availability of COVID-19 therapeutics and needs to be pursued. This approach has not been very successful so far despite several studies published for repurposing; only a couple of drugs (Remdesivir and Baricitinib) have been approved by USFDA for emergency use against COVID-19 (not actually addressing COVID-19 but secondary infections caused by it). Nevertheless, the approach has potential. We have applied our approach for drug repurposing as well with thus far encouraging results [54]. We obtained binding affinities agreeing well with experimental measurements and also gained detailed energetic insight into the nature of drug-protein binding that would be useful in drug discovery for the target studied.

Given the large-scale supercomputing infrastructure available to us, we are able to scale to the vast number of calculations required to provide input to the ML models. Equally important, our methods are designed to provide key uncertainty quantification, a feature vital to our goal of using active learning to optimise campaigns of simulations to maximise the chance that predictive ML models will find promising drug candidates. Our present paper is not a scientific research paper in a conventional sense. We have reported an accelerated drug discovery pipeline but not included any novel scientific findings here, which is going to be the subject of our subsequent publications. Currently, PB components of our workflow have already been implemented successfully in isolation, whereas it is work in progress for ML components which still need some optimisation. Our integrated workflow implementation has also not been fully achieved. In addition, we are working towards improving the computational performance of such a complicated and heterogenous workflow as a whole. We have made substantial progress in this regard in the past few months as described in the following sections. In this paper, we have reported preliminary results obtained using our workflow as it stands now to exhibit that our approach has the potential to impact the process of drug discovery.

## 2. Methods

No single algorithm or method can achieve the necessary accuracy with required efficiency to sample the huge chemical space inhabited by lead compounds for drug discovery. We innovated by combining multiple algorithms into a single unified pipeline (**Figure 1**), using an interactive and iterative methodology, allowing both upstream and downstream feedback to overcome the limitations of classical *in silico* drug design as described above.

We first describe the different components of our workflow, notably their standalone strengths and weaknesses, then show how we couple them constructively in the workflow such that the sum is greater than the parts.

**a) High-throughput Docking**

Protein-ligand docking involves ligand 3D structure (conformer) enumeration, exhaustive docking and scoring, and pose scoring. The input requires a protein structure with a designed binding region, or a crystallized ligand from which a region can be inferred, as well as a database of small molecules to dock, where the chemical structure is represented in the SMILES format.

Conversion of the 2D structures into 3D structures ready for structural docking is performed through proteinisation and conformer generation using Omega-Tautomers that also includes enumeration of enantiomers prior to conformer generation if stereochemistry is not specified [56]. Conformer generation is performed on the ensemble of structures, typically generating 200-800 3D conformers for every enantiomer.

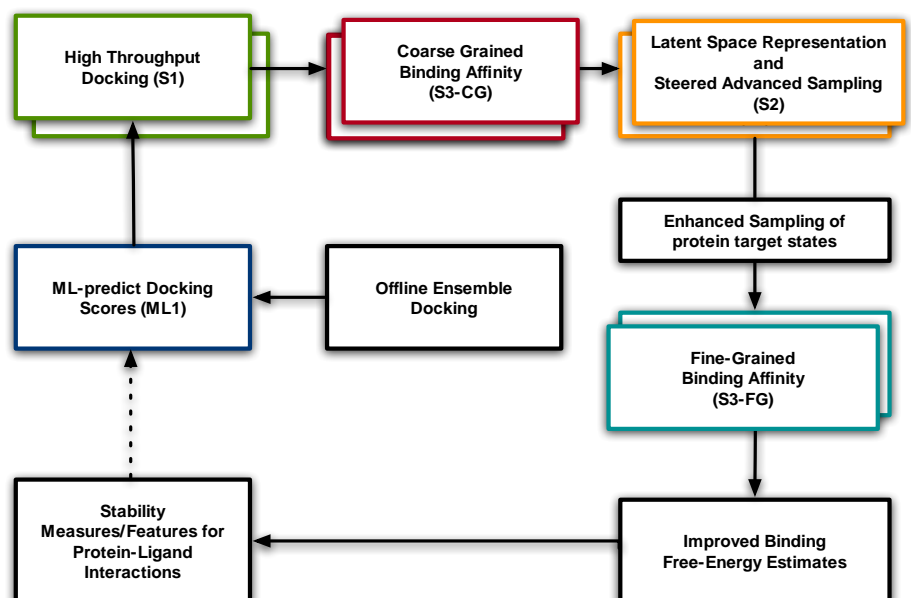

*Figure 1.* Integrated modelling pipeline for new COVID treatments where blind machine learning is made "smarter" with accurate physics based methods. It represents an entire virtual drug discovery pipeline, from hit to lead through to lead optimization. The constituent components are deep-learning based surrogate model for docking (ML1), Autodock-GPU (S1), coarse and fine-grained binding free energies (S3-CG and S3-FG) and S2 (DeepDriveMD). The arrows show the information transferred between the different methods. (Source: IMPECCABLE [55])

Each of the 3D structures so generated is docked against the protein binding pocket and scored. The best scoring pose is returned along with its ChemGauss4 score from exhaustive rigid docking [57]. The ranking obtained using such docking scores are useful in the initial hit identification stage of the drug discovery pipeline.

As a consequence, the outputs of docking runs include a 3D protein (receptor) structure with the docked ligand in its binding site. The docking score (evaluated by the scoring function specific to a docking protocol) provides a qualitative measure of the intrinsic complementarity between a given ligand and protein binding site. While docking protocols are generally good at estimating the binding poses (i.e., 3D conformation) of ligands within a binding site, the energetics of interaction can be challenging to determine and are a function of how a specific scoring function is implemented. Nevertheless, docking is extensively used in structure-based drug design approaches. This is so because docking can predict whether or not a molecule binds at all with the target protein. In addition, given that it is a computationally cheap technique, it makes best economic sense to have an additional filter before performing the expensive binding affinity calculations. In our protocol, docking is implemented at the initial stage to identify area of interest in the chemical space and filter out all the obvious non-binders. Thereafter, we employ MD-based binding affinity prediction methods for more accurate ranking of the available compounds on the top ranked compounds based on their docking scores.

Furthermore, there is a need to account for the intrinsic flexibility of the protein in response to the ligand (which may also induce conformational changes) that needs to be accounted for in the energetics of how ligands/proteins interact. For this purpose, extensive conformational sampling is often necessary. The enhanced/adaptive sampling techniques described below can help addressing some intrinsic limitations of these techniques.

**b) ML-based Conformation Transition Classifier**

In order to investigate the conformational transitions during MD simulations, we used two 10 μs trajectories, made available by D.E. Shaw Research [58], of the SARS-CoV-2 spike glycoprotein starting from two main different conformations (i.e., 6VYB and 6VXX, partially open and closed states, respectively). The dictionary of secondary structure of proteins (DSSP) [59] is used to classify each residue according to its secondary structure in all the frames of the trajectory. A total of 8334 frames are extracted from the 10 μs simulations of the spike glycoprotein. The data used for the analysis consist of the atomic coordinates of the protein's $C^{\alpha}$ atoms and secondary structures of the protein residues, according to DSSP. To analyse the conformations, we adapted the ML-based Anomaly Detection techniques previously designed and employed at the European Organization for Nuclear Research (CERN) for scientific and medical linear accelerators [60]. We predicted the probability of a local protein conformational change based on transitions occurring in individual trajectories.

The trajectory of each $C^{\alpha}$ in class-space is followed in time until a change of class is observed at time $t_a$, with a running over the number of atoms in the simulation. From that time on, the transitions between different classes, if any, are tracked for 100 subsequent frames, forming a corresponding set of stochastic transitions matrices, whose elements $T_{kl}$ represent the transition frequency from class $k$ to class $l$, where $k,l=\{0..7\}$ (cf. Table 1). Only a few transitions out of the possible 64 are effectively observed within the examined dataset. The most frequently observed are the transitions between identical or structurally adjacent classes.

**Table 1:** Classes of secondary structure that DSSP defines.

| Letter ID | Number ID | Class of the secondary structure |
|---|---|---|
| G | 0 | $3_{10}$ helix (first helix) |
| H | 1 | α helix |
| I | 2 | π helix |
| E | 3 | β sheet |
| B | 4 | β bridge |
| T | 5 | helix turn |
| S | 6 | bend |
| C | 7 | coil (No SS found) |

The stochastic transition matrices are then turned into heat maps and fed into a Convolutional Neural Network (CNN). The neural network was a 2-layer CNN, trained using the Reptile meta-learning algorithm [61]. The input layer has a single channel of 8 by 8 pixels. It uses Keras implementation of the *relu* activation function, the Sparse Categorical Cross Entropy Loss Function and the Adam optimizer. The transition-based classification is used to predict the probability of belonging to a class, and of the class that the selected residue might land at a future time, typically after 1500 frames since the initial class change. We compared the prediction with the frequency of belonging to each class, as observed throughout the simulation that was not used for training, i.e. that starting from the 6VXX structure. The similarity between the different distributions was evaluated via the Jensen-Shannon divergence [62]. Our preliminary results, shown in Table 2, are encouraging, although subject to a number of caveats. First and foremost, the training and the validation data set (70:30) pertain to a single trajectory, which implies that some transitions are trained on a very small number of events. Hence, multi-trajectory data are needed to consolidate these preliminary results. Using more data would also allow to introduce additional classes, thus obtaining a more precise estimation of residues' behaviour. This is currently the subject of an on-going research.

**Table 2:** Jensen-Shannon divergence (a measure of the similarity between two probability distributions. bounded between 0 and 1) between predicted and observed secondary class transitions in the 6VXX trajectory of the spike protein system. Data are presented in decreasing order of similarity. The labels code for the initial and final class. When they are identical, it means that after some oscillation, the residue goes back to the initial class.

| Transition | J-S divergence |
|---|---|
| "43" | 4.91E-03 |
| "34" | 5.67E-03 |
| "01" | 7.70E-03 |
| "33" | 1.02E-02 |
| "12" | 1.21E-02 |
| "00" | 1.62E-02 |
| "11" | 1.93E-02 |
| "**21**" | 4.38E-02 |
| "**22**" | 6.14E-02 |
| "**44**" | 6.51E-02 |
| "**10**" | 1.54E-01 |
| "**04**" | 3.68E-01 |

c) **ML-driven Enhanced Sampling**

DL methods have been widely applied to understand protein conformational dynamics, and a number of methods have been proposed to enhance sampling of conformational landscapes using adaptive sampling strategies that include DL methods in their workflows. One such approach, namely DeepDriveMD, uses variational autoencoders to cluster high dimensional data on conformations from multiple MD trajectories into a more manageable low dimensional manifold from which 'novel' conformations can be selected, based on certain reaction coordinates (RCs) or collective variables (CVs) and new simulations can be instantiated from such conformations [63]. This approach has been demonstrated for protein folding trajectories, offering at least 2x speedup compared to traditional conformational sampling methods, and in a recent application, DeepDriveMD was able to enhance sampling by nearly 25% with just 12% of compute time for studying conformational transitions of the SARS-CoV-2 Spike protein bound to the ACE2 receptor [64]. Thus, DeepDriveMD offers a way forward in sampling conformational events, providing a framework to extend its functionality to account for studying protein-ligand interactions.

Ligands bound to the protein target of interest induce specific conformational changes; some ligands may induce changes that are local to the binding site, whereas others may induce changes farther away from the binding site. We posit that even with reasonably short time-scale simulations, our variational autoencoder can cluster protein-ligand interaction landscapes based on such conformational differences and provide a quantitative way to extract ligand-specific protein conformational signatures that could help bound the uncertainty in binding affinity calculations. To this effect, we extracted the contact maps between the protein $C^{\alpha}$-atoms (defined at an 8 Å cut-off) and analysed them with our variational autoencoder. Optimal hyperparameters were determined as described previously [65] and the resulting latent space embedding was visualized using t-Stochastic neighbourhood embedding (t-SNE) approach. In our analysis, we observe clear separation between protein-ligand complexes as well as the fact that some ligands induce more conformational changes than others.

d) **MD-based Binding Affinity Prediction**

Hit-to-Lead (H2L) is a step in the drug discovery process where promising lead compounds are identified from initial hits – small molecules which have the desired activity – generated during preceding stages. After evaluation of initial hits, optimization of promising compounds is carried out to achieve nanomolar affinities. The change in free energy between free and bound states of protein and ligand, also known as binding affinity, is a promising measure of the binding potency of a molecule and is used as a parameter for evaluating and optimising hits at the H2L stage.

A protocol known as ESMACS [13,17] was used to estimate binding affinities of protein-ligand complexes. It involves performing an ensemble of MD simulations followed by free energy estimation using a semi-empirical method called molecular mechanics Poisson-Boltzmann surface area (MMPBSA). The free energies for the ensemble of conformations are analysed in a statistically robust manner, yielding precise free energy predictions for any given complex.

The use of ensembles is particularly important because the usual practice of performing MMPBSA calculations on conformations generated using a single MD simulation does not give reliable binding affinities [66]. Consequently, ESMACS predictions can be used to rank a large number of hits based on their binding affinities. ESMACS is able to handle large variations in ligand structures and hence is very suitable for H2L stage where hits have been picked out after covering a substantial region of chemical space.

The ensemble of conformations for the protein-ligand complex generated using MD simulations are also analysed using the variational autoencoder technique described above to get insights into favourable as well as unfavourable interactions of different functional groups in a molecule with the target protein. This knowledge is helpful in performing further optimization of the lead structures. The information and data generated with ESMACS is additionally used to train our ML model (described below) to improve its predictive capability.

Lead Optimization (LO) is the final step of pre-clinical drug discovery. It involves altering the structures of selected lead compounds in order to improve properties, such as selectivity, potency and pharmacokinetic parameters. Binding affinity is a useful parameter to make *in silico* predictions about the effects of any chemical alteration in a lead molecule. However, LO requires theoretically more accurate (without much/any approximations) methods to make predictions with high confidence. In addition, relative binding affinity of pairs of compounds which are structurally similar are of interest, rendering ESMACS unsuitable for LO.

Because of these issues, we employ thermodynamic integration with enhanced sampling (TIES) [19–21], which is based on an alchemical free energy method called thermodynamic integration (TI) [67]. Alchemical methods involve calculating free energy along a non-physical thermodynamic pathway to get relative free energy between the two end-points. A best practice guide for alchemical free energy calculations was recently published with useful recommendations [68]. Usually, the alchemical pathway corresponds to transformation of one chemical species into another defined with a coupling parameter ($\lambda$), ranging between 0 and 1. TIES involves performing an ensemble simulation at each $\lambda$ value to generate the ensemble of conformations to be used for calculating relative free energy. It also involves performing a robust error analysis to yield relative binding affinities with statistically meaningful error bars. The parameters such as the size of the ensemble and the length of simulations are determined keeping in mind the desired level of precision in the results [19].

e) **ML-based Model to Predict Useful Ligands**

In our drug discovery workflow, ML is used to gather and accumulate information from all the other PB components described above so as to quickly locate the most interesting region(s) in the chemical space in terms of the potential of a lead compound to bind strongly. We have created a ML surrogate model using a simple featurisation method, namely 2D image depictions, as it does not require complicated architectures such as graph convolution networks, while demonstrating good prediction. We obtain these image depictions from the nCov-Group Data Repository [69] that contains various descriptors for 4.2 billion molecules generated on HPC systems with Parsl [70]. By using 2D images, we are able to initialise our models with pre-trained weights that are typically scale and rotation invariant under image classification. This model is used to generate ligand molecules that can be analysed using the PB methods described above. We train our ML model using data from both docking as well as MD based binding affinity predictions so as to enable it to actively relate structural/chemical features with corresponding binding potencies. This allows our ML model to progressively make more accurate predictions of ligand structures that can be classified as initial hits. The predicted structures are then fed into the PB pipeline to filter them, first using docking and then by ESMACS and TIES, to finally select those that bind most effectively. This is repeated with the ML model getting better after each iteration. Thus, we provide reliable training data to our ML models whereas potentially good initial structures are identified for our PB methods. In this way, our workflow couples ML and MD such that each compensates for the weaknesses in the other method. It is our expectation that, together, they are more effective.

## 3. Workflow management

Our workflow (**Figure 1**) integrates different methods and dynamically selects active ligands for progressively computationally expensive methods. At each stage, only the most promising candidates advance to the next stage, yielding a pipeline in which each downstream stage is computationally more expensive, but also more accurate, than upstream stages. Execution of such a complicated workflow requires scalable tools with advanced resource management, task-placement and adaptive execution capabilities, in this case RADICAL-Cybertools (RCT) [71] middleware.

RCT executes tasks concurrently or sequentially, depending on their arbitrary priority relation. Tasks are grouped into stages and stages into pipelines, depending on the priority relation among tasks. Tasks without reciprocal priority relation can be grouped into the same stage, tasks that need to be executed before other tasks have to be grouped into different stages. Stages are then grouped into pipelines and, in turn, multiple pipelines can be executed either concurrently or sequentially. RCT uses RADICAL-Pilot (RP) [72] to execute tasks on HPC resources, allowing the execution of workflows with heterogeneous tasks.

RCT middleware has been used in two ways:

*Scalable concurrent multi-stage task-execution:* A work around was required to use the middleware on one of Europe's largest supercomputers, SuperMUC-NG. As RADICAL-Cybertools depend on 3rd party software modules, the virtual environment required by RP couldn't be created on SuperMUC-NG because access is granted only to allowed IP addresses. Thus, we prepared RCT's virtual environment outside of the system and then moved it to SuperMUC-NG login node. In this way, RCT could be launched from a login node via the pre-set environment, without the need for outbound internet access. RCT uses MongoDB and RabbitMQ as communication services. These services need to be accessible from both login and compute nodes. On SuperMUC-NG, we automated the launching of both services on a dedicated compute node, which was provided by a special service queue with unlimited walltime, while the workers for RP were provided by the regular batch system.

*Concurrent Multiple Workflow Execution:* RCT's fine-level task-placement feature allows the concurrent use of both CPUs and GPUs on supercomputer nodes. That is achieved by employing RADICAL-Pilot's unique capability of concurrently executing heterogeneous tasks on CPU cores and GPUs as an integrated hybrid workflow. This allows the concurrent execution and interleaving of different workflows, making better use of compute resources. RP places tasks on specific compute nodes, cores and GPU [73]. When scheduling tasks that require different amounts of cores and/or GPUs, RP keeps tracks of the available slots on each compute node of its pilot. Depending on availability, RP schedules CPU tasks (e.g., MPI) within and across compute nodes and reserves a CPU core for each GPU task. This results in efficient placement of heterogeneous tasks on heterogeneous resources.

Leveraging aforementioned RP's heterogeneous task placement capabilities, we merge ESMACS and TIES into an integrated hybrid workflow with heterogeneous tasks that utilise CPU and GPU concurrently. Running these two calculations concurrently reduces the total execution time, substantially saving computational cost, thereby improving resource utilisation at scale.

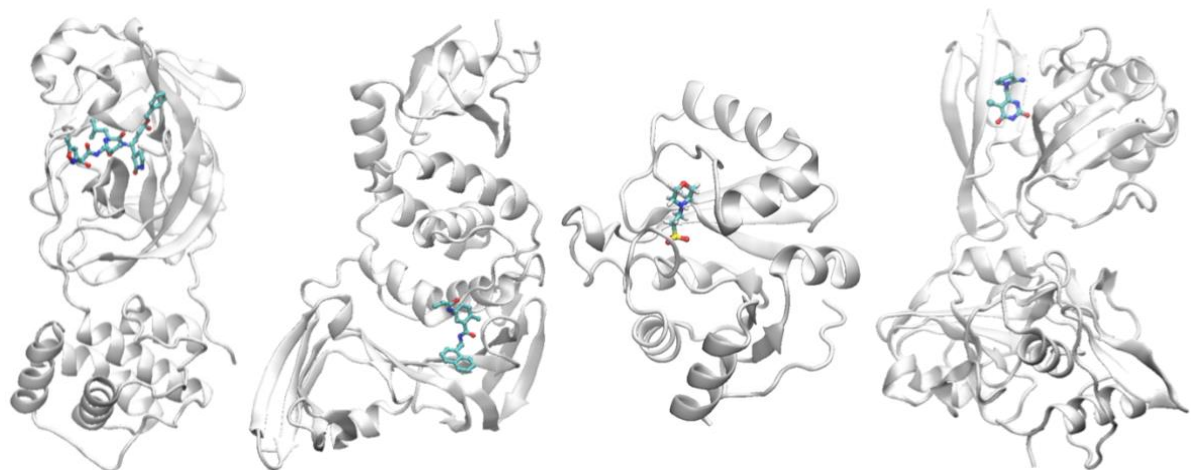

*Figure 2. Structures of the four target proteins studied, in each case shown bound to a compound. From left to right: 3CLPro, PLPro, ADRP and NSP15. The proteins are shown in cartoon representation, and compounds in stick representation.*

In the past few months, we have progressed substantially with the implementation of our workflow. RCT is now fully functional on Summit [55, 73-74] as well as Theta [75], in addition to several other HPC resources. It has successfully executed workloads at 95% usage on these machines. We characterized scaling performance of various components of our workflow using up to 392,000 cores and 24,582 GPUs to execute 24,552 heterogeneous executable tasks and $126 \times 10^6$ python function tasks [74]. Recently, we have been able to achieve a performance of 144M/hour docking hits screening ~$10^{11}$ ligands using over 8000 compute nodes, which is better than the previous best by a factor of two [76]. This has substantially boosted our ability to screen large compound libraries as well as generating training data for surrogate models. We have already analyzed several million compounds from a set of orderable compound libraries using the current implementation of our workflow and filtered out compounds for the second iteration of our iterative workflow. Recently accepted publications in IEEE TPDS, ACM SIGHPC ICPP, ACM SIGHPC PASC [55, 73-74], as well as publications under review [75, 76] provide evidence of our progress towards the fully optimised implementation of the workflow.

**4. High performance computing resources**

Our workflow is by design based on high throughput computational (HTC) calculations. Even though, it reduces the overall number of necessary computations tremendously, an acceptable time to solution is only achievable on HPC resources. To illustrate the impact of our workflow, we applied it to four target proteins of SARS-CoV-2 in this work, namely 3C like protease (3CLPro; also known as the main protease), papain-like protease (PLPro), ADP-ribose phosphatase (ADRP; a macrodomain of NSP3), and non-structural protein 15 (NSP15) (**Figure 2**). These proteins have diverse functions for the replication and transcription of the coronavirus, and are important targets for pharmaceutical drug design and discovery [77-81]. For this, docking calculations were performed on thousands of ML-model-generated ligand conformations, leading to a ranking of candidates with corresponding ligand structures. Afterwards, we conducted several hundred ESMACS calculations on the top ranked ligands based on their docking and 19 TIES calculations on a selection of ligand pairs. Note that the ML-based generation of ligand structures accelerated the whole HTC process significantly.

We would like to emphasise here that the above results are only preliminary and do not constitute novel scientific findings. The above mentioned calculations were performed at a small scale for testing and optimizing our workflow. We have the PB components already working well in isolation. However, we are still optimizing the DeepDriveMD protocol for application in prospective drug discovery. In addition, we are yet to realise the fully optimised implementation of our workflow as a whole with all its components working in tandem. This paper is about the development of the infrastructure, so we have not included novel scientific results in terms of potential drug candidates identified in this paper. Nevertheless, we have started applying the current implementation of our workflow to a large-scale dataset of millions of orderable compounds. Using our docking protocols, we identified 10,000 compounds for each of the 4 target proteins which were used for performing ESMACS calculations. The top 500 compounds based on their ESMACS ranking are being further optimized using DDMD (as it stands) to identify potentially better binding conformations that will be used for the second iteration of our workflow. This work is underway, and we have some very encouraging results with input from experimental colleagues that will be published in due course.

Drug repurposing is another promising approach that bypasses all the stringent requirements of drug approval, and hence could accelerate the availability of COVID-19 therapeutics. We have recently used our workflow to make a detailed assessment of a set of proposed repurposed drugs [54]. We obtained binding affinities agreeing well with experimental measurements and also gained detailed energetic insight into the nature of drug-protein binding that would be useful in drug discovery for the target studied.

All calculations were performed on a variety of supercomputers including Leibniz Rechenzentrum's SuperMUC-NG, Hartree Centre's ScafellPike, Oak Ridge National Laboratory's Summit and Texas Advanced Computing Center's Frontera. Table 2 summarizes performance and cost numbers for the calculations on Summit to understand the overall cost of the presented pipeline. Note that the ESMACS calculations were accelerated with OpenMM as MD engine on GPUs. TIES required longer wall-clock times as only CPUs were employed to obtain the data for Table 2. However, recently we have developed a GPU-enabled version of TIES on Summit (using NAMD3 as well as OpenMM as MD engines) which costs only 11 node-hours per ligand-protein complex. This would substantially reduce the computational cost associated with our workflow.

**Table 3:** Overview of computing cost for the different calculations in the computing pipeline on Oak Ridge National Laboratory's Summit supercomputer.

| Calculation | Physical time required in each MD simulation [ns] | # of independent MD simulations per ligand-protein complex | Computing time per calculation [node-hours] | Computing time per ligand-protein complex [node-hours] | Used theoretical node performance [TF] |
|---|---|---|---|---|---|
| **Docking** | --- | Several thousands | 0.0001 | --- | --- |
| **ESMACS** | 12 | 25 | --- | 10 | 420 |
| **TIES** | 6 | 65 | --- | 700 | 29400 |

## 5. ESMACS and TIES applied to COVID-19 on high end resources

### a) ESMACS findings

ESMACS is used at the hit identification and H2L stage of the drug discovery. The DL-based surrogate model was used to screen the small molecules in the Zinc database, a collection of commercially available chemical compounds. A high-throughput docking study was then performed to generate binding poses of the compounds to the 4 COVID-19 proteins (**Figure 2**). While docking programs are generally good at pose prediction, they are less effective in predicting binding free energy of the compounds to the target proteins. To better rank the binding potentials of the compounds, we performed ESMACS simulations for the top 100 compounds for each of the selected proteins. The compounds were chosen from 10,000 docked small molecules, based on their docking scores from the high throughput docking study.

**Table 4:** Number of the most promising compounds for each of the 4 proteins investigated. For each protein, the top 100 compounds, chosen from 10,000 docked small molecules, are evaluated by ESMACS approach. The number of compounds are listed, which have the most favourable binding free energies, in the ranges corresponding to $K_D$ values of 10 nM (-10.98 kcal/mol), 100 nM (-9.61 kcal/mol), 1 μM (-8.24 kcal/mol).

| Energy (kcal/mol) | 3CLPro | ADRP | NSP15 | PLPro |
|---|---|---|---|---|
| ΔG < -10.98 | 1 | 0 | 3 | 6 |
| $-10.98 \leq \Delta G < -9.61$ | 2 | 2 | 1 | 8 |
| $-9.61 \leq \Delta G < -8.24$ | 1 | 4 | 10 | 5 |
| $\Delta G < -8.24$ Total | 4 | 6 | 14 | 19 |

Preparation and setup of the simulations were implemented using binding affinity calculator (BAC), including parametrization of the compounds, building simulation-ready topologies and structures of the complexes, and generating configurations files for the simulations. MD simulations were performed using two MD engines, NAMD and OpenMM, on three machines, Frontera, Summit and SuperMUC-NG. For each replica, energy minimisations were first performed, followed by 2 ns equilibration. Finally, 10 ns production simulations were run for each replica. MMPBSA calculations were then performed for all of the 1000 frames from the 10 ns production runs, while configurational entropies were calculated using NMODE on 48 or 56 frames for each replica, depending on the number of cores per node on the computers used for NMODE calculations.

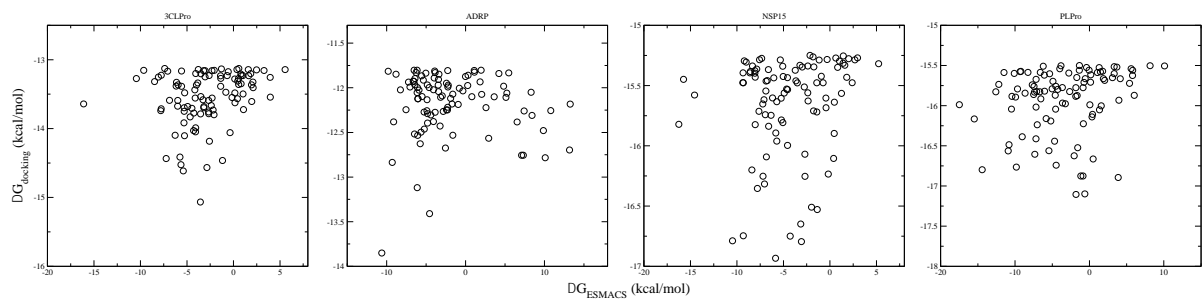

*Figure 3. Correlations between the ESMACS results and docking predictions. Weak or no correlations are obtained for the 4 protein targets – 3CLPro, ADRP, NSP15 and PLPro – with correlation coefficients of 0.25, -0.06, 0.16 and 0.20, respectively.*

For most of the molecular systems studied, about 4-19% of the compounds show promising binding affinities (cf. Table 3), with free energies more negative than -8.24 kcal/mol (corresponding to a $K_D$ value on the nanomolar scale). Although the distributions of predicted free energies from independent simulations are non-normal [24,25], the ensemble-based ESMACS predictions are highly reproducible, independent of which MD code is used, or on which supercomputer the simulations are performed. As stated above, the docking scores are not a good indicator for binding affinities; the free energies from ESMACS calculations only show weak correlations with the docking scores (**Figure 3**). The inclusion of configurational entropy has a negligible impact on the ranking of the binding free energies. The ESMACS study shows that the most promising compounds can be selected more reliably using the ESMACS prediction than the docking scores.

b) **TIES findings**

TIES is used at the LO stage of drug discovery to hone interactions between protein and ligand so to enhance the binding potency of selected lead compounds. To demonstrate this capability, we performed TIES on a set of 19 compound transformations (that is chemically mutating the "original" compound into a "new" compound) to study the effect of small structural changes on a compound's binding potency with ADRP. The calculated free energy differences show a non-normal nature, as we have recently reported [24,25]. The ensemble-based TIES approach ensures high precision predictions, with uncertainties less than 0.82 kcal/mol for all but one of the calculations (cf. Table 4). The relative binding affinities (ΔΔ$G$) predicted by TIES for these transformations fall between -0.55 to +4.62 kcal/mol (cf. Table 4). A positive value indicates a diminished relative binding potency for the "new" compound, whereas a negative value means that the transformation studied is favourable. 12 out of the 19 transformations studied have ΔΔ$G$ > +1 suggesting that they all correspond to unfavourable structural changes. The remaining seven transformations have statistically zero value for ΔΔ$G$, which implies that the corresponding structural modifications do not affect the binding. It is difficult to predict what structural changes will improve the binding. Thus, the knowledge of both "useful" as well as "rubbish" transformations is of much value at the LO stage so as to make informed structural changes. TIES provides us an excellent tool to do so with confidence. Such information then informs our ML predictive model about the desirable as well as undesirable chemical modifications to be introduced into the selected lead compounds. In this way, it improves the predictive accuracy of the ML models, progressively leading to quicker convergence towards the region of our interest in the huge chemical space.

**Table 5:** Results from TIES calculations on a set of ligand transformations studied for ADRP. ΔΔ***G*** is the relative binding affinity for a transformation, that is the change in binding affinity on morphing one ligand into the other. ***σ*** corresponds to the uncertainty associated with the relative binding affinity predicted by TIES.

| Transformation | ΔΔ***G*** (kcal/mol) | ***σ*** (kcal/mol) |
|---|---|---|
| a0-a2 | 1.48 | 0.60 |
| a0-a4 | 1.82 | 0.66 |
| a0-a5 | 1.14 | 0.60 |
| a0-a6 | 3.22 | 0.44 |
| a0-a7 | 1.32 | 0.43 |
| a0-a9 | 0.25 | 0.57 |
| a0-a10 | 1.52 | 0.70 |
| a0-a41 | 3.41 | 0.53 |
| a0-a44 | 1.18 | 0.49 |
| a0-a45 | -0.46 | 0.52 |
| a0-a46 | 2.91 | 0.70 |
| a0-a47 | 0.36 | 0.57 |
| a0-a48 | -0.55 | 0.57 |
| a0-a49 | 1.84 | 0.46 |
| a0-a50 | 0.52 | 0.64 |
| a1-a42 | -0.29 | 0.82 |
| a1-a43 | 2.05 | 1.03 |
| a3-a42 | 0.49 | 0.81 |
| a42-a43 | 4.62 | 0.82 |

6. **Conclusions**

We describe an innovative, iterative and interactive heterogenous workflow that has the potential to accelerate the existing drug discovery process substantially by coupling machine learning with physics-based methods such that each compensates for the weaknesses of the other. This workflow requires high-throughput screening of a large number of small molecules based on their binding potencies evaluated using various types of methods. Molecules filtered at one stage are advanced to the next to be filtered once again using a more accurate and computationally intensive method. A refined set of lead compounds emerges at the end of this multistage process for wet lab studies. With information relating structural features to energetics and binding potencies being fed into the ML model at each stage, it learns how to improve the prediction of the next set of compounds. This iterative process, along with the upstream and downstream flow of information, allows it to accelerate the sampling of relevant chemical space much faster than traditional methods. We have demonstrated the application of our workflow on four SARS-CoV-2 target proteins. The workflow requires HPC resources for efficient implementation and a dedicated workflow manager to handle the large number of heterogenous computational tasks on a multitude of supercomputers. We believe that this hybrid ML-PB approach offers the potential in the long term – with the rise of exascale, quantum and analogue processing – to deliver novel pandemic drugs at pandemic speed.

## Data accessibility

The models and simulation trajectories were generated at UCL. Models used for performing PB simulations and results obtained are available at the following public github repository: https://github.com/UCL-CCS/ML-PB-Covid-drug . Docking related codes are available at https://github.com/inspiremd/Model-generation, whereas ML related codes and sample files are located on https://github.com/inspiremd/molecular-active-learning. Sample scripts for executing our workflow using RCT are also available at https://github.com/inspiremd/Model-generation.

## Authors' contributions

APB, SW and DA performed PB simulations, data curation and analysis. ARC, TB, AP, FX, XD, AF, HM, AR, WR, NS, SS, SV, YD and ADM performed ML related calculations and analyses. LT, MiT, AM, MaT and SJ provided workflow management infrastructure. DA, SJ, DK, AR and PVC acquired computational resources for the current study; GM and DW provided technical support. PVC organised and led the overall project. APB, SW, RH and PVC composed the paper with substantial input from all authors. All authors contributed to the editing and reviewing of the paper, and read and approved the manuscript.

## Funding

We are grateful for funding for the UK MRC Medical Bioinformatics project (grant no. MR/L016311/1), the UK Consortium on Mesoscale Engineering Sciences (UKCOMES grant no. EP/L00030X/1) and the European Commission for the EU H2020 CompBioMed2 Centre of Excellence (grant no. 823712), the EU H2020 EXDCI-2 project (grant no. 800957), as well as financial support from the UCL Provost. Our research was also supported by the DOE Office of Science through the National Virtual Biotechnology Laboratory, a consortium of DOE national laboratories focused on response to COVID-19, with funding provided by the Coronavirus CARES Act. This research was supported as part of the CANDLE project by the Exascale Computing Project (17-SC- 20-SC), a collaborative effort of the U.S. Department of Energy Office of Science and the National Nuclear Security Administration. The work has been supported in part by the Joint Design of Advanced Computing Solutions for Cancer (JDACS4C) program established by the U.S. Department of Energy (DOE) and the National Cancer Institute (NCI) of the National Institutes of Health. Anda Trifan acknowledges support from the United States Department of Energy through the Computational Sciences Graduate Fellowship (DOE CSGF) under grant number: DE-SC0019323.

**Competing interests**

The authors declare no competing financial interest.

**Acknowledgements**

Access to SuperMUC-NG, at the Leibniz Supercomputing Centre in Garching, was made possible by a special COVID-19 allocation award (award ID COVID-19-SNG1) from the Gauss Centre for Supercomputing in Germany. We acknowledge excellent support from Don Maxwell, Bronson Messier and Sean Wilkinson at OLCF. We also wish to thank Dan Stanzione and Jon Cazes at Texas Advanced Computing Center. Some of this work was performed thanks to a 2021 DOE INCITE award "COMPBIO".